# Optical modulators with two-dimensional layered materials

Zhipei Sun[1*], Amos Martinez[2*], Feng Wang[3*]

[1] Department of Micro- and Nanosciences, Aalto University, Finland
[2] Aston Institute of Photonic Technologies, Aston University, UK
[3] Department of Physics, University of California, Berkeley, California 94720, USA

*E-mail: zhipei.sun@aalto.fi; a.martinez3@aston.ac.uk; fengwang76@berkeley.edu

**Abstract:** Light modulation is an essential operation in photonics and optoelectronics. With existing and emerging technologies increasingly demanding compact, efficient, fast and broadband optical modulators, high-performance light modulation solutions are becoming indispensable. The recent realization that two-dimensional layered materials could modulate light with superior performance has prompted intense research and significant advances, paving the way for realistic applications. In this review, we cover the state-of-the-art of optical modulators based on two-dimensional layered materials including graphene, transition metal dichalcogenides and black phosphorus. We discuss recent advances employing hybrid structures, such as two-dimensional heterostructures, plasmonic structures, and silicon/fibre integrated structures. We also take a look at future perspectives and discuss the potential of yet relatively unexplored mechanisms such as magneto-optic and acousto-optic modulation.

**Introduction**

Optical modulation is one the most crucial operations in photonics. It is ubiquitous in photonic and optoelectronic applications, such as optical interconnects, environmental monitoring, bio-sensing, medicine or security applications. We are amidst the era of information where internet applications (e.g., media streaming, cloud computing, internet of things) continue to develop at an extremely fast pace. This is generating an exponentially increase in the number of network data interconnections, which include traditional data network and intra-/interchip data connections. The dominant electronic interconnection approach (e.g., copper cables) suffers from issues of bandwidth and loss due to its performance restrictions in terms of speed, energy consumption, dispersion and cross-talking. This leads to an urgent need for alternative short-reach interconnect methods with better performance. Optical solutions offer intrinsic advantages in terms of higher-bandwidth and lower-loss. Therefore, intense research efforts are being directed towards light modulation aiming to develop compact, cost-effective, efficient, fast and broadband optical modulators for high-performance optical interconnects[1]. This will also greatly impact other applications, such as fibre-to-the-home, environmental monitoring, astronomical, bio-sensing, and medical applications[1].

In recent years, graphene and other two-dimensional (2D) layered materials (e.g., transition metal dichalcogenides (TMDs) and black phosphorus) are attracting increased attention for applications in electronics, photonics and optoelectronics[2], due to their unusual electrical and optical properties. 2D layered materials can exhibit a rich variety of physical behavior, ranging from that of a wideband insulator to a narrow-gap semiconductor, to a semimetal or metal. They provide exciting opportunities for diverse photonic and optoelectronic functions enabling new conceptual photonic devices, fundamentally different from those based on traditional bulk materials[2-4]. For example, graphene, the best known 2D material, has been widely used for numerous photonic and optoelectronic devices, operating at an extremely broad spectral range extending from the ultraviolet, visible and near-IR to the mid-IR, far-IR, and even to the THz and microwave regions due to graphene's unique linear energy-momentum dispersion relation[2-4]. The demonstrated devices include transparent electrodes in displays, solar cells, optical modulators and photodetectors[2-4].

Aside from graphene, monolayer TMDs (such as $MoS_2$[5,6], and $WS_2$) and black phosphorus[7] are direct bandgap semiconductors, offering properties complementary to graphene. Different atomically-thin 2D materials can be readily stacked together by van der Waals force to make 2D heterostructures without the conventional "lattice mismatch" issue, offering a flexible and easy approach to design desired physical properties. The surfaces of 2D materials are free from any dangling bonds[2,4], compatible with different photonic structures, such as well-developed fibre[8] and silicon[9,10] devices, potentially for large-scale and low-cost integration into the current dominant optical fibre network and silicon CMOS technology.

Compared with traditional bulk semiconductors, 2D materials also provide additional values, such as mechanical flexibility, easy fabrication and integration, and robustness[2]. Furthermore, previous demonstrations of graphene and other 2D materials suggest that almost all functions required for integrated photonic circuit (e.g., generation, modulation, and detection of photons) can be accomplished by 2D materials[2]. Such versatility of operation combined with their unique electronic properties (i.e., high mobility, bandgap tunability), may enable the integration of 2D material based electronic and photonic devices together to achieve multifunction integrated photonic and electronic circuits.

Optical modulation effects in 2D layered materials are among the most extensively studied research topics over the last few years. Prominently, this leads to massive prototype demonstrations on optical modulators with different modulation mechanisms (e.g., all-optical, electro-optic, thermo-optic modulations), showing competitive performance. For example, the most distinctive performance in reported graphene optical modulators is its extremely broad operation bandwidth, which in total can cover from the visible to the microwave regions. These demonstrated 2D material based optical modulators have already been effectively used in numerous applications. For instance, graphene, other 2D materials and their heterostructure based saturable absorbers, an all-optical passive modulator, have been successfully implemented for ultrafast pulse generation in a variety of lasers, demonstrating superior performance[8,11]. This greatly encourages current commercialization interest for various laser applications, including high-repetition-rate ultrafast laser sources for optical interconnections[8,11]. Graphene enabled nonlinear wavelength modulators[12] and electro-optic amplitude modulators[9,10] also have been tested for high-speed data transmission experiments at speeds up to 22 Gbit/s[10], promising for high-speed light modulation in optical interconnections.

In this Review, we present the state-of-the-art of optical modulators based on 2D materials including graphene, TMDs and black phosphorus. We will also cover recent advances employing their hybrid structures such as 2D heterostructures, plasmonic modulators, and silicon/fibre integrated modulator devices. Finally, we will conclude with a compressive discussion on prospective of 2D layered materials for future applications.

> **Box 1: Fundamentals of optical modulators**. Optical modulators can be categorized by different ways. For example, depending on the attribute of light that is modulated (Boxfig.1a), optical modulators can be classified into amplitude modulators, phase modulators, polarization modulators, wavelength modulators, etc. Depending on the principle of operation (Boxfig. 1b), optical modulators can be classified into all-optical, electro-optic, thermo-optic, magneto-optic, acousto-optic and mechano-optic modulators, etc. Depending on the optical property of the material changed for light modulation, primary modulators can be categorized as absorptive modulators or refractive modulators. For absorptive modulators, the absorption coefficient of the material (i.e., the imaginary part of its refractive index) is controlled by an absorption related effect, such as saturable absorption, electro-absorption, Franz-Keldysh effect and quantum-confined Stark effect. In refractive modulators, light modulation is generally realized by effects correlated with the change of the real part of refractive index, such as the Kerr effect, Pockels effect, thermal modulation of the refractive index, change of the refractive index with sound waves (i.e., acoustic waves).

The key figures of merit used to characterize a modulator are modulation speed, modulation depth, operation wavelength range, energy consumption and insertion loss[1]. Modulation speed is a critical parameter for optical modulators. It is typically defined by the operation frequency when modulation is reduced to half of its maximum value. Fast modulation is generally needed for most data transmission applications (e.g., network interconnects), where the modulation speed is normally measured by its ability to modulate optical signals at a certain data transmission rate (e.g., bit/s). The modulation depth is often measured by the extinction ratio, i.e., the ratio between the maximum and minimum transmittance ($T_{max}/T_{min}$) in transmission-type devices (or the ratio between the maximum and minimum reflectance $R_{max}/R_{min}$ in reflection-type devices). Decibel unit is widely used for its simplicity in engineering. As a result, the modulation depth is typically given by $10\times\log(T_{max}/T_{min})$ (or $10\times\log(R_{max}/R_{min})$). High modulation depth (>7 dB) is preferable for most applications (e.g., high data-rate interconnects, high-sensitivity sensing), however, < ~4 dB modulation depth can be adequate for certain applications (e.g., passive mode-locking, short-distance data transmission)[1]. Another important parameter of a modulator is its operation wavelength range. For optical data transmission systems, modulators are typically required to operate at one or more of the three major telecom windows (i.e., ~ 0.85, 1.3 and 1.5 μm), or at the visible range for emerging visible light wireless (Li-Fi) communications. For some applications such as data centres and high-performance computing, energy efficiency is now a critical requirement. The targeted energy consumption for future energy-efficient optical modulators is estimated to be around a few fJ/bit for on-chip connections (about two orders of magnitude below the current power consumption level)[13]. Optical devices and systems, including modulators, have the potential to outperform their electronic counterparts in terms of lower energy consumption and higher data connection speed. Insertion loss of a modulator is also of practical significance as it directly relates to the system energy efficiency. In addition, other considerations, such as stability (e.g., thermo-stability), compatibility (e.g., waveguide, Boxfig. 1b), footprint and cost, are also essential towards evaluating any potential trade-offs between the performance metrics discussed above for practical applications.

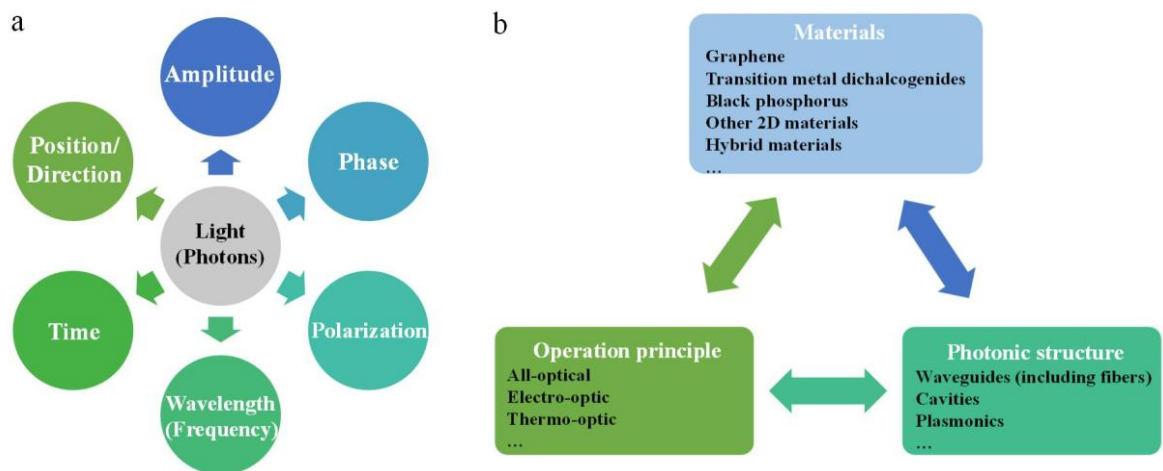

**Boxfigure 1 a**, Attributes of light that can be modulated mainly include amplitude, phase, polarization, wavelength (frequency), time (for pulsed optical beam), and position (direction). **b,** Key considerations for optical modulators, involving with material, operation principle, and photonic structure design.

## Fundamental electronic and optical properties of 2D layered materials

In 2D layered materials with single-unit-cell thickness, new electronic and optical properties can emerge due to quantum confinement, enhanced electron-electron interactions, reduced symmetry, and proximity effects. In addition, unprecedented control of these properties can be achieved through a large range of methods for light modulation, such as electrostatic gating, mechanical strain, and substrate engineering in 2D layered materials. Next we briefly review the unique fundamental electronic and optical properties of the most widely studied 2D materials, including graphene, TMDs, black phosphorus, as well as layered heterostructures.

**Graphene.** Graphene, a single atomic layer of hexagonal carbon lattice, has been extensively studied due to its unique mechanical, thermal, electronic, and optical properties. It has an electronic structure characterized by two linear Dirac cones at the K and K' points at the Brillouin zone[14] (Fig. 1a). Electrons in graphene behave as massless Dirac fermions with forbidden backscattering, and exhibit the highest mobility at room temperature[15]. Photons across a wide range of the electromagnetic spectrum interact strongly with graphene. Optically, the linear dispersion of electrons lead to a strong and universal absorption of $\pi\alpha$ in pristine graphene across the whole infrared to visible spectral range, where $\alpha = e^2/\hbar c$ is the fine-structure constant[16,17]. Remarkably, the strong and broadband light-matter interaction in graphene can be controlled effectively by tuning the Fermi energy $E_F$ of graphene with electrostatic gating. On the one hand, interband transitions with energy below $2E_F$ become forbidden due to Pauli blocking. Consequently graphene becomes essentially transparent for photons with energies up to $2E_F$[18,19]. On the other hand, intraband transitions from free carriers increase dramatically upon gating, leading to strong Drude-like absorption peaks in the infrared[20,21]. In addition, this free-carrier response of graphene supports 2D plasmon mode, which exhibits unusually strong confinement and a distinctive dependence on carrier concentration[22,23]. The ability to control broadband graphene absorption and plasmon excitation through electrostatic gating enabled many electro-optic modulator designs based on graphene that function at the terahertz to visible wavelengths[9,22,24,25].

**Transition metal dichalcogenides.** Layered TMDs, in which the d-orbital electron interactions lead to a rich variety of physical properties ranging from semiconductors to charge density waves and to superconductors, offer a rich platform to explore novel 2D phenomena. Semiconducting TMDs (such as $MoS_2$, $MoSe_2$, $WS_2$, $WSe_2$) are particularly interesting for optoelectronic applications. These semiconducting TMDs are indirect semiconductors in bulk. However, they exhibit an indirect- to direct-bandgap transition when thinned to monolayers[5,6] (Fig. 1b), with a direct bandgap ranging from 1.57 to 2 eV for different TMDs[26,27]. As a result, photoluminescence in monolayer TMDs can be orders of magnitude stronger than that in their bulk counterparts although the amount of material in monolayer TMD is much less[5,6]. Optical absorption in monolayer semiconducting TMDs is remarkably strong, reaching over 10% at bandgap resonances[5]. Theoretical and experimental studies show that these optical resonances are dominated by excitonic transitions in semiconductor TMDs, where the exciton binding energy can be at hundreds of meV due to dramatically enhanced electron-electron interactions in 2D monolayers[28-31]. In addition, a pair of degenerate exciton transitions are present at the K and K' valley in momentum space of TMD monolayers with broken inversion symmetry, giving rise to a unique valley degree of freedom that is analogous to electron spin[32]. Polarization-resolved photoluminescence studies show that the valley pseudospin in TMDs can couple directly to the helicity of excitation photons, raising the intriguing prospect of valleytronics that exploits the valley degree of freedom[32]. Electrically, field effect transistors with high on-off ratio have been realized in TMD monolayers, thanks to the large semiconductor bandgap[33]. The electron mobility in TMDs is relatively low, typically limited to 0.1-100 $cm^2/V\cdot s$ at room temperature[34].

**Black phosphorus.** Black phosphorus is another single-element layered material. Monolayer and few-layer phosphorene are predicted to bridge bandgap range from 0.3 to 1.5 eV[35], between the zero bandgap of graphene and bandgaps higher than 1.57 eV in semiconducting TMDs (Fig. 1c). Inside monolayer phosphorene, each phosphorus atom is covalently bonded with three adjacent phosphorus atoms to form a puckered, honeycomb structure. The three bonds take up all three valence electrons of phosphorus, so unlike graphene, monolayer phosphorene is a semiconductor with a predicted direct optical bandgap of ~ 1.5 eV at the Γ point of the Brillouin zone. The bandgap in few-layer phosphorene can be strongly modified by interlayer interactions, which leads to a bandgap that decreases with phosphorene thickness, eventually reaching 0.3 eV in the bulk limit. Unlike TMDs, the bandgap in few-layer phosphorene remains direct for all sample thickness. This makes black phosphorus an attractive material for mid-IR and near-IR optoelectronics. The puckered structure of black phosphorus breaks the three fold rotational symmetry of a flat honeycomb lattice, and leads to anisotropic electronic and optical properties. It was shown that optical absorption and photoluminescence in black phosphorus are highly anisotropic[7,36-38]. Field effect transistors based on few-layer black phosphorus exhibit reasonably high on-off ratio of $10^5$ at room temperature, which is between the values achieved in graphene and TMD based transistors[39]. Electron mobility of few-layer phosphorus is highly anisotropic, reaching 1500 cm$^2$/V.s along x-direction and 800 cm$^2$/V.s along y-direction at low temperature[36]. This mobility is significantly higher than that in TMDs. One material challenge facing black phosphorus is its air sensitivity: monolayer and few-layer phosphorus degrade quickly in ambient condition, and hematic sealing will be required for any black phosphorus based devices.

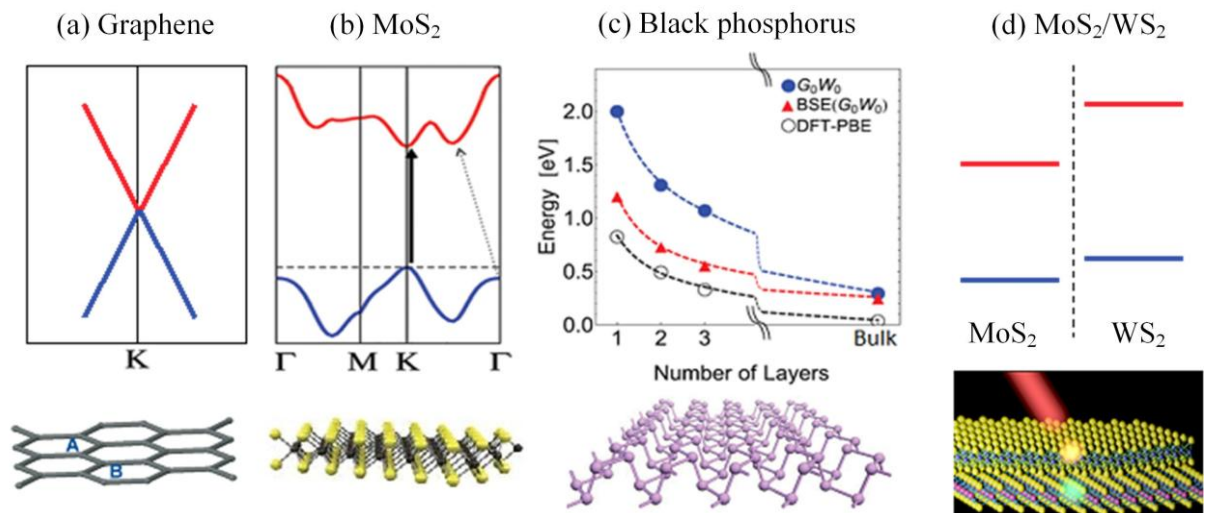

**Figure 1.** Electronic structure of different 2D layered materials. **a,** Graphene is a zero-bandgap semiconductor with massless Dirac electrons and a linear Dirac cone at the K point. **b,** Monolayer semiconductor TMDs such as MoS$_2$ has a direct bandgap at the K point, in contrast to the indirect bandgap in their bulk counterparts[6]. **c,** Few-layer black phosphorus is predicted to be direct-bandgap semiconductors with a bandgap energy ranging from 0.3-1.5 eV[35]. **d,** Layered heterostructures can have unusual band alignments, such as the type-II semiconductor junctions between MoS$_2$ and WS$_2$ that facilitate efficient charge transfer processes[40].

**Layered material heterostructures.** Atomically thin 2D layers with wide-ranging properties can be prepared separately and then stacked together to form van der Waals-bonded heterostructures, in which each layer can be engineered separately. There have been tremendous efforts to explore different 2D heterostructures, including but not limited to graphene/*h*BN, TMD/*h*BN, TMD/graphene, and TMD/TMD combinations. The graphene/*h*BN heterostructures are characterized by fascinating Moire superlattice physics that gives rise to mini- Dirac points[41] and Hofstadter butterfly pattern[42-44]. Heterostructures including TMDs are particularly exciting for optoelectronic and light harvesting applications because many TMD monolayers are direct-bandgap semiconductors with remarkably strong light-matter interactions. For example, TMD/graphene heterostructures enabled novel memory devices[45] and ultrathin photodetectors[46,47]. TMD/TMD heterostructures with different 2D materials, on the other hand, form type-II heterojunctions (Fig. 1d). Such TMD/TMD heterostructures allow efficient separation of photoexcited electrons and holes, where ultrafast charge transfer between MoS$_2$/WS$_2$ at femtosecond time scale has been observed[40,48]. In addition, atomically thin PN junctions have been realized based on MoS$_2$/WSe$_2$ heterostructures[49]. The freedom of combining a rich variety of different materials in van der Waals heterostructures is likely to lead to even more exciting discoveries of novel electronic and optical properties.

**State-of-the-art of optical modulators with 2D materials**

Research on optical modulators with 2D materials has attracted huge interest, and this has translated into tremendous progress over the past few years. In this section, we review the state-of-the-art of various optical modulators based on 2D materials, including all-optical modulators, electro-optic modulators, thermo-optic modulators and other less-explored modulators.

**All-optical modulators.** All-optical light modulation using 2D layered materials has been extensively studied since it allows for the signal processing to be realised fully in the photonic domain. Thus, modulation can be done directly in an optical fibre or other waveguide (e.g., silicon waveguide) system, allowing low-loss and broadband optical signal processing in simple configurations. Demonstrated all-optical modulators with 2D materials include saturable absorbers[50-52], wavelength convertors[53], optical limiters[54], and polarization controllers[55]. The majority of these devices exploit the strong nonlinear optical response of 2D materials (mainly the third order susceptibility), their broad bandwidth, fast response and miniature size for compact, integrated all-optical operation. For example, the imaginary part of the third order nonlinearity Im($\chi(3)$) is responsible for saturable absorption, a mechanism employed for passive mode-locking and Q-switching of lasers. The real part, Re($\chi(3)$) is responsible for nonlinear processes such as four-wave mixing and third harmonic generation.

Driven by the growing need of ultrafast lasers, 2D material based saturable absorbers are amongst the earliest and most successful photonic devices utilizing 2D materials. Here, the 2D material operates as a passive self-amplitude modulator that enables ultrafast pulse generation[50-52,56,57]. As discussed, graphene's band structure ensures the existence of electron-hole pair excitations in resonance with any incoming photon from the visible to the far-IR. The interaction between charge carriers and ultrafast optical pulses produces a non-equilibrium carrier population in the valence and conduction bands, which relaxes on an ultrafast time scale[3]. This guarantees wideband and ultrafast saturable absorption from Pauli blocking. In practise, however, a relatively large saturation fluence at wavelengths shorter than the near-IR spectral region has hindered graphene's applicability at that end of the spectrum[8]. Unlike graphene, TMDs[5,6,26,27] and black phosphorus[35] exhibit finite bandgaps for resonant light absorption. For example, TMDs[56-58] typically have resonant absorption in the visible, and black phosphorus[38,59] shows resonant absorption in the near-IR and mid-IR. This offers a suitable alternative to graphene saturable absorbers at those wavelengths. For example, Ref. [60] reported TMD-based (i.e., WS$_2$, MoS$_2$, and MoSe$_2$) saturable absorbers for all-fibre pulsed lasers in the visible regime (Fig. 2a), opening a new way for next-generation high-performance pulsed laser sources in the visible (even ultraviolet) range. An additional interesting feature of TMDs is that they can work as saturable absorbers at wavelengths below the bandgap[57,58] possibly due to sub-bandgap absorption from crystallographic defects and edge states[58] in 2D TMD flakes[61].

Intensive research effort on 2D material saturable absorbers, particularly for ultrafast pulse generation, has led to a steady performance improvement[8,11]. For example, most laser formats (e.g., fibre (Fig. 2a,b), waveguide, solid-state, and semiconductor lasers[8,11]) have been successfully modulated for ultrafast pulse generation, delivering pulse repetition rate up to 10 GHz (Fig. 2b)[62], pulse width down to sub-100 fs[11] and with a broad wavelength coverage[8,11] expanding from the visible (e.g., 635 nm[60]) to the mid-IR (e.g., 2.9 μm[8]) range. External cavity optical processing (e.g., amplification, frequency conversion and pulse compression) also has been demonstrated as a convenient way to push the output performance[8,11], e.g., the wavelength accessibility, the pulse duration and the output power and pulse energy. A particularly innovative approach is currently undergoing development in order to improve the ultrafast laser performance with a combined active and passive modulation function in 2D materials[63-66]. Figure 2c proposes a high-performance ultrafast laser enabled with 2D material based photonic integrated circuit. Such concept potentially could lead to compact, reliable, low-noise and cost-effective ultrafast lasers, by which various emergent applications (e.g., in telecommunication and metrology) will be underpinned.

Borrowing the concept of pump-probe spectroscopy that is widely used to study carrier dynamics in nanomaterials (including 2D materials), all-optical active modulation has been realized by exploiting different mechanisms (e.g., Pauli blocking[67] and optical doping[68]). In this case, light transmission through 2D materials at the signal's wavelength is modulated (or switched) by another light beam with photon energy higher than the signal[67,68]. Ref. [67] demonstrated an all-optical modulator with a single-mode microfibre wrapped with graphene (Fig. 2d), achieving 38% modulation depth and ~2.2 ps response time, which is limited by ultrafast carrier lifetime in graphene. This approach is in principle suitable for ultrafast signal processing with modulation rate of >200 GHz[67]. In free-space setups, wide-band (from ~ 0.2 to 2 THz) THz light modulation has been demonstrated with a maximum modulation depth of 99% in a graphene-silicon structure by exploiting the optical doping effect[68].

Experimental studies on the coherent nonlinear optical response of graphene have confirmed its very large third order susceptibility using four-wave mixing[53] and, more recently, using third harmonic generation[69]. This shows the possibility of nonlinear optics based wavelength modulators for ultrafast all-optical information processing (e.g., all-optical wavelength conversion[53]) due to fast response (typically in fs region) of third-order nonlinear susceptibility. An early study demonstrated wavelength conversion of a 10 Gbit/s non-return-to-zero signal in an all-fibre configuration using graphene deposited on a fibre-end (Fig. 2e)[12]. The device had a conversion efficiency of -27 dB and a wavelength detuning of 12 nm. However, various previous studies on graphene and other 2D materials also evidenced that, in order to fully harness the potential of the strong nonlinearity of 2D materials at the atomic scale, it is necessary to circumvent the issues of inefficient light-matter interaction due to 2D material's sub-nanometer thickness as well as optical damage due to the high optical excitation power required. Several approaches (e.g., stacking multiple monolayer[8], evanescent mode integration, doping[70], interference effect[71], coherent control[72], microcavity (Fig. 2f)[73], and slow-light waveguide[74]) have demonstrated the possibility of enhancing light-matter nonlinear optical interaction in 2D materials in recent years. For example, by placing graphene in a high Q-factor silicon photonic crystal microcavity (Fig. 2f), resonant optical bi-stability, temporal regenerative oscillations and cavity enhanced four-wave mixing in graphene have been demonstrated at light intensities as low as a few fJ[73]. 2D material based heterostructures (e.g., $MoS_2$-$WS_2$[40,48], $MoS_2$/$WSe_2$[48,49]) have also been proposed for novel linear and nonlinear optical device designs with tunable optical properties (e.g., reflectance[48], carrier dynamic[40]), but full potential of 2D heterostructures for nonlinear optics still deserves further exploration.

Up to now, most practical all-optical photonic devices using 2D materials rely on third order nonlinear processes. Graphene is centrosymmetric, so it only exhibits weak second harmonic generation[75]. Other 2D materials with broken symmetry, however, can exhibit a strong second order nonlinearity. This has been demonstrated on several 2D materials (e.g., $MoS_2$[61,76,77], BN[77], $WS_2$[78], $WSe_2$[78,79]), where samples with an odd number of layers exhibits second-order nonlinearity orders of magnitude higher than that with an even number of layers. The strength of the generated second harmonic can be electrically controlled with exciton related effects[79], and is strongly dependent on the crystallographic orientation of the crystal[61]. Consequently, second harmonic generation has proven very useful for the characterization of the number of layers and orientation[61] of TMDs. Furthermore, nonlinear frequency conversion process (including four-wave mixing and the reverse process of harmonic generation) is the dominant method to achieve quantum light sources. High optical nonlinearity in 2D materials (e.g., graphene, TMDs, black phosphorus) may make high-purity quantum emitters and quantum optical switches[80] possible for integrated quantum circuits with atomic thickness.

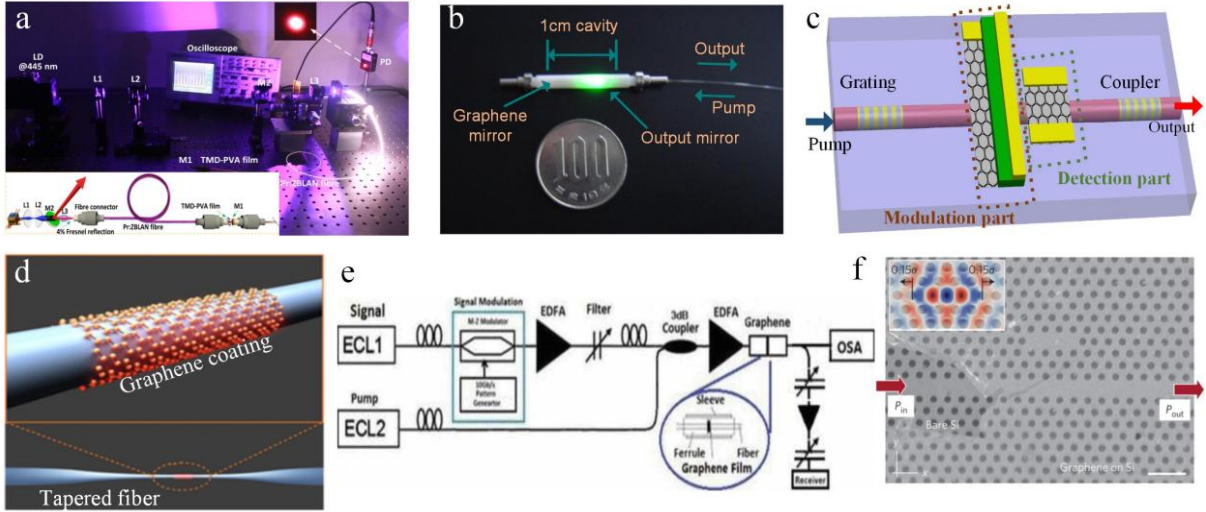

**Figure 2. 2D material based all-optical modulators. a,** Photograph of a TMD-based saturable absorber modulated visible fibre laser (Inset: schematic of the laser setup)[60]. **b,** Photograph of a high-repetition-rate graphene mode-locked fibre laser[62]. **c,** Schematic of a high-performance ultrafast laser enabled with 2D material based photonic integrated circuit. This can be realized by integrating a modulator and photodetector into a waveguide or fibre laser system. The modulation part can provide active/passive mode-locking and laser stabilization, while the detection part can provide feedback for laser stabilization. **d,** Schematic of a graphene-clad microfibre all-optical modulator[67]. **e,** All-fibre experimental setup on four-wave mixing based wavelength conversion in graphene[12]. **f,** Graphene-clad silicon photonic crystal nanostructure for enhanced nonlinear frequency conversion[73]. Scale bar, 500nm.

**Electro-optic modulators.** Electro-optic modulators exploit electro-optic effects to electrically control the light properties. They are particularly desirable for data communication link applications. Thus far, 2D material based electro-optic modulators have been demonstrated mainly by utilizing the gate-tunable electro-absorption effect in graphene[9,18,19]. Recently, the tunability of the refractive index of graphene has also been experimentally demonstrated by gating[81,82], indicating the possibility of using 2D materials for electro-refractive phase modulators. Other electro-optic effects in 2D materials, such as Franz-Keldysh effect[2] and quantum-confined Stark effect[2], are also possible for light modulation, but have not yet been experimentally demonstrated.

Early results on 2D material based electro-optic modulators revealed the significant advantages of using 2D materials owing to their broad operation bandwidth, compactness, low operation voltage, ultrafast modulation speed, and CMOS compatibility[9]. In contrast to the conventional semiconductor materials, where absorption is limited by their bandgap, graphene absorbs light across a broad electromagnetic spectral range from the UV to the THz. This enables light modulation with a much broader operation wavelength range. Indeed, light modulation with graphene has been demonstrated covering the visible[25,63], IR[9,63-66,83,84] and THz[21,85-90] range. At the visible and near-IR range, the typical modulation speed of graphene electro-absorption modulators can be on the order of GHz (e.g., 1 GHz[9,83,91], 30 GHz[10]). The intrinsic speed of typical electro-optic modulators is limited by its RC time constant, which therefore can be reasonably predicted far beyond 100 GHz[10,93,94], comparable to the state of the art of high-speed 2D material electronics[2].

2D materials demonstrate strong light-matter interaction. The absorption coefficient of graphene and monolayer $MoS_2$ is larger than $5\times10^7$ m$^{-1}$ in the visible range, which is an order of magnitude higher than the bandgap absorption of GaAs and Si[4]. However, the absolute value is very small for such atomically thin materials. For example, monolayer graphene absorbs ~2.3% of white light[16,17]. This means that the intrinsic modulation of monolayer graphene based free-space devices can only be up to 2.3% (~0.2 dB). This value is insufficient for most practical applications that typically require signal modulations of at least ~50% (3 dB), for example, telecom applications. Therefore, it is crucial to overcome the low absorption in monolayer 2D materials in order to increase the modulation depth. Various methods have been proposed or demonstrated to improve the modulation depth by using multi-layer devices (including few-layer graphene[25] or stacked monolayer graphene[83] devices), reflection-mode[84], patterned structure[95], interference enhancement[84], evanescent-mode coupling[65] and cavities[96].

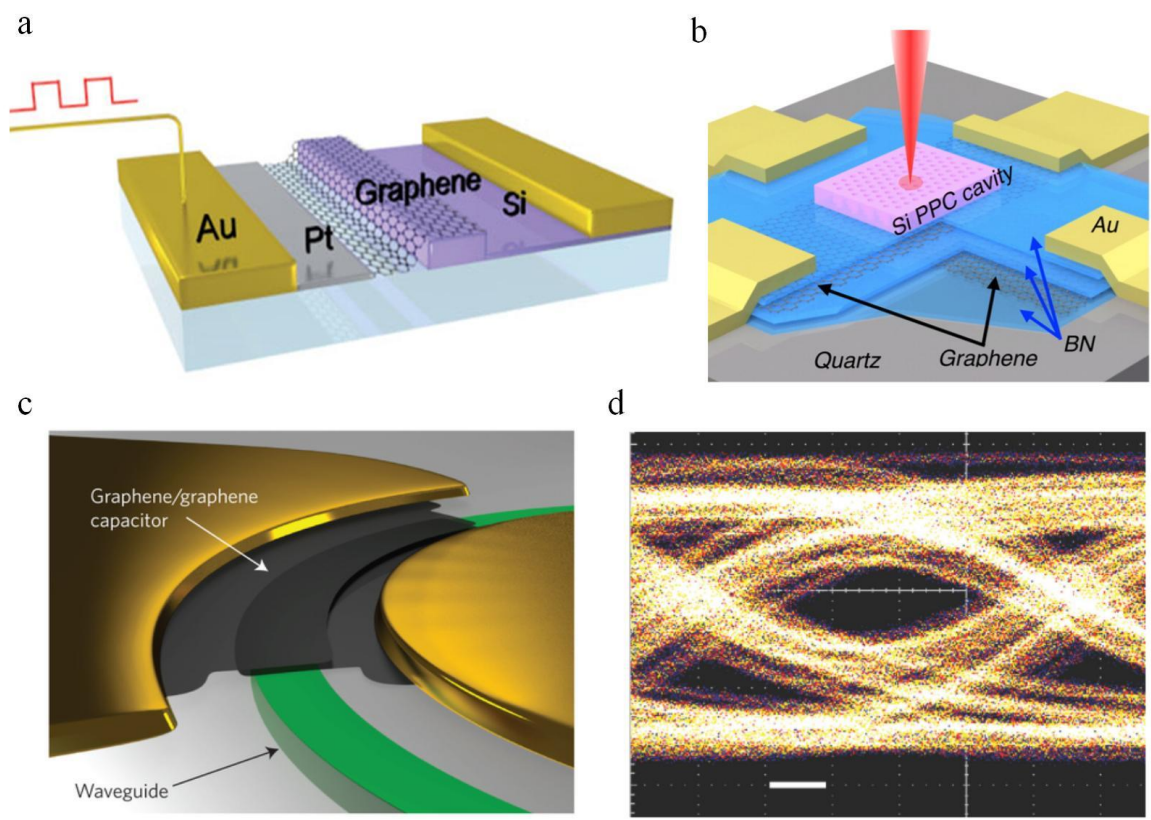

**Figure 3. 2D materials and their heterostructure based electro-optic modulators for silicon photonics. a,** Schematic of the first graphene based silicon waveguide modulator[9]. **b,** Schematic of a graphene-BN heterostructure based photonic crystal cavity modulator[91]. **c,** Schematic of a graphene based silicon nitride ring resonator modulator[10] and **d,** its 22 Gbit/s data transmission experiment[10]. Scale bar, 8ps.

For silicon integrated photonics, the current leading candidate technology for short-reach optical interconnects[1], the issue of the low modulation depth of 2D material based devices has mainly been addressed by using evanescent-mode coupling in different formats, including non-resonance waveguide (Fig. 3a)[9,83,92,97,98], interferometers[99] and cavity enhancement approaches (Fig. 3b,c) [100]. Ref. [91] demonstrated a graphene-BN heterostructure based electro-optic modulator integrated with a silicon photonic crystal cavity (Fig. 3b), achieving operation frequency up to 1.2 GHz with a modulation depth of 3.2 dB. Similarly, silicon micro-ring resonators can provide efficient light modulation with various advantages, such as small footprint, large modulation depth, and small energy consumption, which have also been used for 2D material based silicon modulators[101,102] (Fig. 3c). Ref.[10] demonstrated a graphene electro-optic micro-ring silicon nitride modulator with critical coupling effect, whereby an increase in loss of a coupled micro-ring resonator increases the device transmission by changing the resonance coupling condition. This offers considerable improvement in the performance (e.g., a 30-GHz bandwidth with a modulation efficiency of 15 dB per 10 V)[10]. It must be noted that resonant structure approaches can have drawbacks, such as relatively small operation wavelength range and high sensitivity to fabrication tolerance and temperature variation. To date, experimental demonstrations of 2D material integrated silicon modulators have shown impressive performance (>15 dB modulation depth[10,98], ~hundred fJ/bit energy consumption[10,92,100], 30 GHz bandwidth[10] (Fig. 3d)), already comparable with current semiconductor modulation technologies (e.g., silicon[1]). Further down the line, theoretical calculation predicts that by using high-quality 2D materials and optimizing the device structure[10,93], the performance could reach operation speeds in excess of 100 GHz[93,94] and energy consumption levels below 1 fJ/bit[94].

THz research has been one of the most investigated research fields in the last decade. There is a great demand for components in this spectral range for numerous applications extending from health, environmental applications to security applications, but conventional optical components have been challenged in this spectral range. Graphene modulators working at the THz region (Fig. 4a,b)[21,85-90] have been demonstrated through modulation of the intra-band absorption, giving good modulation performance, such as >93% modulation depth[89,90]. Ref.[103] even extended the operation wavelength to the microwave range at 10.5 GHz frequency (wavelength of 2.8 cm) with graphene integrated active surfaces that can electrically control the reflection, transmission and absorption of microwaves (Fig. 4c,d). In this THz spectral range, the demonstrated modulation speed is on the order of KHz or MHz range[85-90], which is limited by the large size (~mm, comparable to the THz beam waist) of the device.

It has been shown that graphene is a promising plasmonic material for manipulating electromagnetic signals at the deep-subwavelength scale owing to its unique physical properties[23], e.g., high carrier mobility and electrostatically tunable optical properties. Consequently, graphene plasmonic electro-optic modulators have been demonstrated in the THz, and IR range[22,104,105] due to its primitive frequency response. Pattered structures (e.g., ribbons[22], disks[106], and stacked multi-layer designs[106]), hybrid structures (e.g., graphene-gold nanoparticles[107,108], graphene-BN heterostructures[109,110] and plasmonic waveguide[111]) and metamaterials[24] (e.g., Fano resonances[112], Metal-insulator-metal waveguides[113]) have been discussed to tune the device response, which can possibly enable significant nonlinear optical interactions at the few photon level[80]. In particular, metamaterial structures with 2D materials[24,112,113] are exciting considerable attention for light modulation (e.g., the wavelength[106], amplitude[24,112,113], phase[24], and polarization[22,106] modulation), showing significantly improved modulation performance with broad operation bandwidth (e.g., covering from the IR to THz), high modulation depth (e.g., up to 95%[89,113] (Fig. 4e)) and modulation speed (e.g., <10 ns response time in the mid-IR region[113]). Note that, in principle, similar to graphene, other 2D materials (e.g., 2D topological insulators and superconductors) can also be effectively utilized for light modulation with similar strategies.

**Thermo-optic modulators.** Thermo-optic effects are also used for light modulation. The most common type is based on refractive modulation, which uses the change in the material refractive index associated with variations in the temperature. Unsurprisingly, thermo-optic modulators are rather slow (~ MHz) due to the intrinsically slow thermal diffusivity. Therefore, thermo-optic modulators are considered for applications where high-speed is not necessary, such as optical routing and switching.

Because of its high intrinsic thermal conductivity[2], graphene is very attractive for various thermal applications, such as flexible and transparent heaters and conductors[2]. Graphene based electric heaters have been integrated into graphene based long-range surface plasmon waveguides (Fig. 5a)[114] and silicon ring resonators[115] for light modulation by thermally induced refractive index change. Graphene based transparent flexible heat conductors also have been used to deliver localized heat for light modulation in a silicon based Mach-Zehnder interferometer (Fig. 5b) and a micro-disk resonator [116]. Another recent study used an optically-controlled graphene heater to change the fibre's refractive index in a fibre based Mach-Zehnder interferometer for all-fibre phase shifter and switching[117]. Thus far, demonstrated graphene thermo-optic modulators have high extinction ratios as high as 30 dB[114], and MHz[115] response, comparable to the traditional material based thermo-optic modulators. However, owning to its excellent thermal property[2], graphene should be capable of operating at faster modulation speeds (up to tens of MHz[115]). Other 2D layered materials might also be suitable for thermo-optic refractive modulation, in particular when operating at wavelength outside the bandgap due to the low insertion losses. High thermal conductivity in 2D materials combined with the easy fabrication and integration, makes them a suitable and cost-effective solution for applications not requiring ultrafast modulation speed (e.g., short-distance optical communication and sensing).

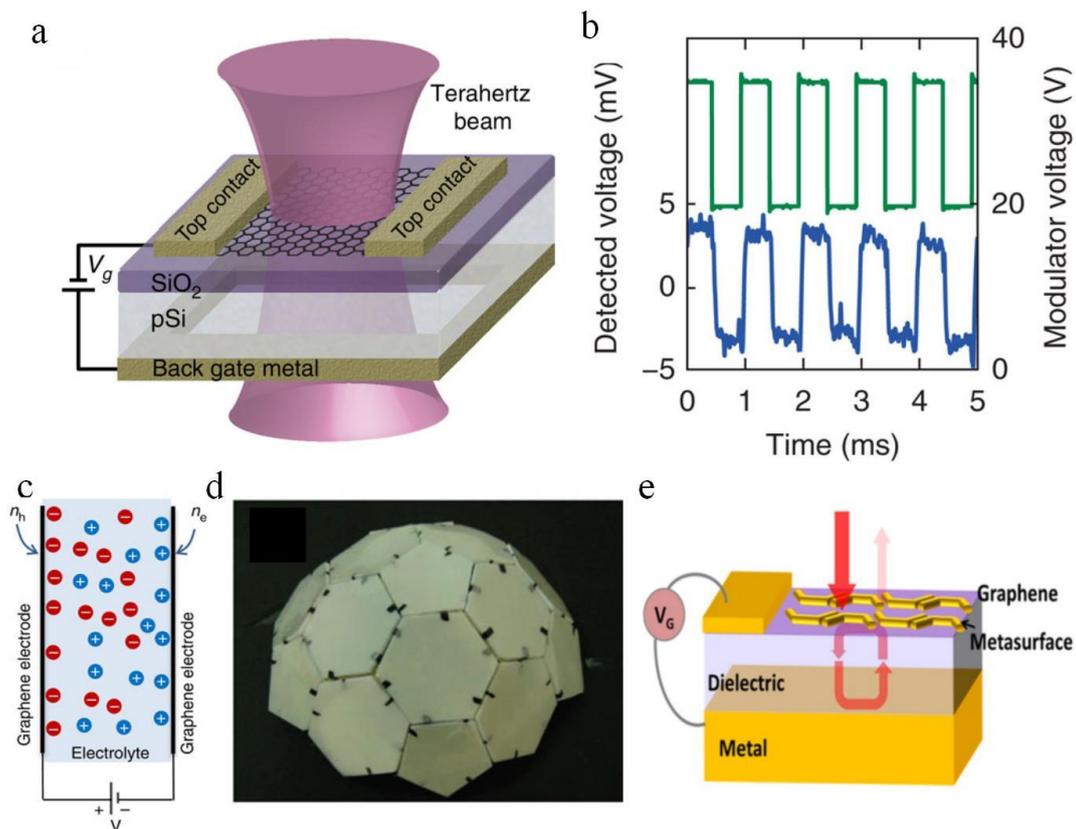

**Figure 4. 2D material based electro-optic modulators at the THz, mid-IR, and microwave range. a,** Schematic of a graphene modulator at the THz range and **b,** its response[85]. **c,** Schematic of a graphene modulator for adaptive microwave surfaces and **d,** its photograph[103]. **e,** Electrically tunable graphene based metasurface perfect absorber in the mid-IR spectral range[113].

**Magneto-optic modulators.** Magneto-optic modulators employing magneto-optic effects (e.g., Faraday effect or magneto-optic Kerr effect) for light modulation are yet to receive as much attention as all-optical or electro-optic modulators. This is partially because of the operation simplicity of the all-optical and electrical approaches. However, a unique nonreciprocal property in magneto-optic modulators offers the opportunity of creating various devices with special functions that are not feasible with other modulators. In particular, magneto-optic modulators may find applications such as optical isolators, circulators, polarization controllers, and electric/magnetic field sensors.

Magneto-optic Faraday (Fig. 5c)[118-120] and Kerr[118] rotation has been experimentally reported in graphene at the far-IR[119], THz[118] and microwave[120] range, indicating the possibility of graphene based magneto-optic modulators for various non-reciprocal applications. A Faraday rotation of up to 0.1 rad (~6°) at the magnetic field of 7 T (Fig. 5d) has been demonstrated in the far-IR range, originating from the excitation of the cyclotron resonance[119]. This is an extremely large Faraday rotation, if we consider that this is achieved with a graphene monolayer with single-atom-thickness, but practical applications will require significant improvements in performance, such as reducing the required magnetic field and pushing its operation towards shorter wavelengths. The magneto-optic response (e.g., Faraday rotation and cyclotron resonance) can be enhanced using cavities, magnetoplasmons[121,122], and metastructures[123]. In other 2D materials such as TMDs and black phosphorus, various magneto-optic effects (e.g., Faraday rotation, Zeeman effect and magneto-optical Kerr effect) have also recently been addressed, offering new opportunities to use these nanomaterials for magneto-optic modulators at the nanoscale for future applications, such as isolators, circulators, and magnetic measurements.

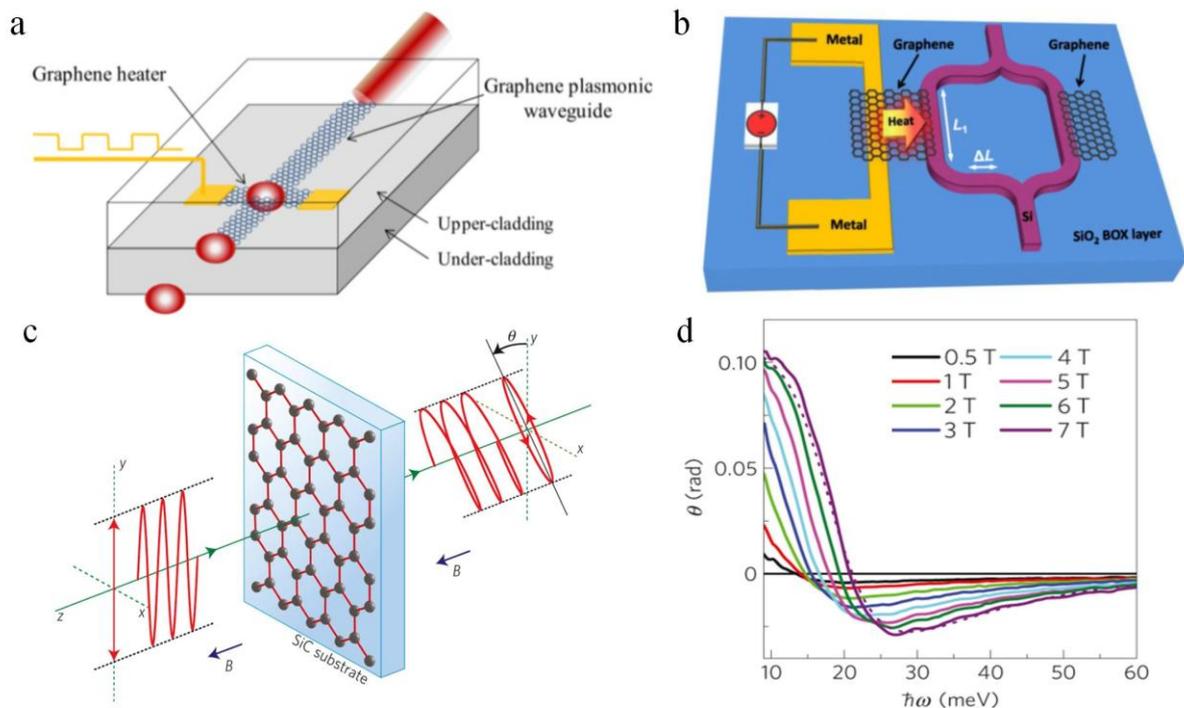

**Figure 5. 2D material based thermo-optic and magneto-optic modulators. a,** Schematic of a thermo-optic modulator based on a graphene heater[114]. Thermally induced inhomogeneous refractive-index distribution of the cladding polymer controllably attenuates the graphene based long-range surface plasmon waveguide for light modulation. **b,** Schematic of a thermo-optic modulator based on graphene heat conductor[116]. **c,** Schematic of a graphene based Faraday-rotator and **d,** its Faraday-rotation response at different magnetic fields[119].

**Acousto-optic modulators.** Acousto-optic modulators mainly use acoustic waves to change the refractive index of a material for light diffraction and frequency changing. Acousto-optic modulators have been widely used for pulse generation (e.g., Q-switching), signal modulation in optical telecommunications and displays. Graphene[124] and other 2D materials (e.g., $MoS_2$[125]) are attracting growing interest from the field of acoustics. These 2D materials are being considered for the generation, propagation, amplification, and detection of surface acoustic waves[124,125]. This gives an indication that it is technically feasible to conceive 2D material based acousto-optic modulators. In addition, periodic diffraction gratings generated in graphene with a sound wave can enable efficient excitation of surface plasmon polaritons[126], demonstrating a simple way to realize plasmonic light modulation with surface acoustic wave. Due to the large surface area and unique properties (e.g., high frequency sensitivity to absorbed molecules), 2D material based acousto-optic modulators can potentially be used to miniaturize the current bulk acousto-optic modulators for specific applications, such as gas sensing and displays.

**Other modulation approaches.** Graphene and other 2D materials have unique mechanical properties including a high Young's module combined with a low mass, suggesting they are a promising material for optomechanics (i.e., mechano-optical modulators). For example, graphene membranes can be actuated up to high mechanical vibration frequencies (> few hundreds MHz), which can be accommodated for modulation of microwave photons[127]. Stress-induced physical property changes in 2D materials can also be used for mechano-optical modulators. In addition to the aforementioned physical methods (e.g., optical, electrical, and thermal methods), which provide simple and effective ways for light modulation with 2D layered materials, other approaches, such as chemical or biological means, can also be utilized to modify the properties of 2D materials for light modulation. This capability makes 2D material based optical modulators potentially for a wide range of applications in biomedical instrumentation, chemical analysis, environmental sensing and surgery.

**Perspectives**

For practical applications, superior performance is always advantageous. In this Review, we have discussed the substantial improvements in the performance of 2D material based optical modulators. For example, the modulation speed of graphene electro-optic modulators has seen a 30-fold improvement over the past 4 years, from 1 GHz[9] to 30 GHz[10]. Nonetheless, there is still a significant need for performance improvement in order to compete with the well-developed traditional technologies, for example on modulation speed and energy consumption for short-reach optical interconnection applications.

2D materials can enable various functions[2] (e.g., electronics, energy storage and conversion, sensor, photonic and optoelectronic functions) due to their diverse physical properties. For example, multifunctional optical modulators (e.g., multifunctional modulator and photodetector[99], multifunctional modulator and plasmon waveguide[114]) have been demonstrated with 2D materials. This not only offers a significant benefit for electronics and photonics in terms of integration with an "all-in-one" solution, but also enables new devices with superior performance (e.g., simultaneous modulation and detection[99]). However, the trade-off between performance and cost might make the commercial success of all-2D material systems challenging in the short term. A more realistic success for 2D materials in the short term would build on the well-developed photonic technology platforms of silicon photonics and fibre optics, where the bulk of research on 2D material based modulators is conducted. In this case, excellent integration capabilities and competitive performance have been achieved showing that 2D materials are ready to become a complementary technology to empower the traditional photonic platforms (e.g., fibre optics and silicon photonics).

Further, intense research on 2D material based modulators has already revealed several distinct advantages of 2D material modulation technology (e.g., broad operation bandwidth from the visible to THz range, small-footprint, low-cost, and large flexibility). And those characteristics have led to an enormous range of new photonic devices, such as, tunable notch filters[106], spatial light modulators[88], smart windows[3], controllable photonic memories[24], tunable adaptive cloaking[103], just to name a few. Given the extremely fast pace at which the 2D materials are being studied, the large variety of available 2D materials, their heterostructures and hybrid systems may continue to introduce new modulation concepts based on their unique physical properties, and more importantly, performance improvements to outperform competing technologies. Therefore, if production of large-scale and high-quality 2D materials is successful[2], it is anticipated that 2D material based optical modulators might create a completely new market to address ongoing challenges and emerging applications (e.g., visible wireless communication, mid-IR/THz bio-sensing and pharmaceutical applications, wearable and bendable photonic applications), potentially revolutionizing the current photonics.

**Acknowledgments**

ZS acknowledges funding from the European Union's Seventh Framework Programme (REA grant agreement No. 631610), Teknologiateollisuus TT-100, Academy of Finland (No.:276376, 284548), and Nokia foundation. AM acknowledges funding from the H2020 Marie-Sklodowska-Curie Individual Fellowship scheme. FW acknowledges funding from the United States National Science Foundation (EFMA-1542741).